\documentstyle[11pt]{article}

      \makeatletter
      
      \@addtoreset{equation}{section}
      \makeatother

\newcommand{\emp}[1]{{\bf #1}}

\newtheorem{teiri}{\emp{Theorem}}[section]
\newtheorem{kei}{\emp{Corollary}}[section]
\newtheorem{meidai}{\emp{Proposition}}[section]
\newtheorem{hodai}{\emp{Lemma}}[section]
\newtheorem{teigi}{\emp{Definition}}[section]
\newtheorem{rei}{\emp{Example}}[section]
\newtheorem{chui}{\emp{Remark}}[section]

\newcommand{\proof}{\noindent {\em Proof:\ } \ \ }
\newcommand{\QED}{\hspace*{\fill}$\Box$ \\}

\newcommand{\chuiowari}{\hspace*{\fill}$\Box$ \\}

\newcommand{\Prob}[1]{{\rm Pr}\left\{#1\right\}}

\newcommand{\defarrow}{\stackrel{\rm def}{\Longleftrightarrow}}

\newcommand{\spr}[1]{{\bf #1}}
\newcommand{\pliminf}{\mbox{p-}\liminf}
\newcommand{\plimsup}{\mbox{p-}\limsup}

\newcommand{\vep}{\varepsilon}
\newcommand{\vph}{\varphi}

\newcommand{\barH}{\overline{H}}

\newcommand{\ubarI}{\underline{I}}

\newcommand{\dss}{\displaystyle}

\newcommand{\cB}{{\cal B}}

\newcommand{\cD}{{\cal D}}

\newcommand{\cM}{{\cal M}}

\newcommand{\cV}{{\cal V}}
\newcommand{\cX}{{\cal X}}
\newcommand{\cY}{{\cal Y}}
\newcommand{\cZ}{{\cal Z}}

\newcommand{\sV}{\spr{V}}
\newcommand{\sX}{\spr{X}}
\newcommand{\sY}{\spr{Y}}

\newcommand{\sW}{\spr{W}}
\newcommand{\sbold}[1]{\mbox{{\scriptsize\bf #1}}}

\newcommand{\ssv}{\spr{v}} 
\newcommand{\ssx}{\spr{x}}

\newcommand{\ssy}{\spr{y}}

\newcommand{\uI}{\ubarI}

\newcommand{\nth}{\frac{1}{n}}
\newcommand{\nti}{n \to \infty}

\newcommand{\noi}{_{n=1}^{\infty}}
\newcommand{\lsn}{\limsup_{n \to \infty}}
\newcommand{\lin}{\liminf_{n \to \infty}}
\newcommand{\lmn}{\lim_{n \to \infty}}

\newcommand{\uIXY}{\ubarI(\sX;\sY)}

\newcommand{\pyn}{P_{Y^n}}

\newcommand{\MID}{\nth \log \frac{W^n(Y^n|X^n)}{P_{Y^n}(Y^n)}}

\newcommand{\bteiri}{\begin{teiri}}
\newcommand{\eteiri}{\end{teiri}}
\newcommand{\bkei}{\begin{kei}}
\newcommand{\ekei}{\end{kei}}
\newcommand{\brei}{\begin{rei}}
\newcommand{\erei}{\end{rei}}
\newcommand{\bhodai}{\begin{hodai}}
\newcommand{\ehodai}{\end{hodai}}
\newcommand{\bteigi}{\begin{teigi}}
\newcommand{\eteigi}{\end{teigi}}
\newcommand{\bchui}{\begin{chui}}
\newcommand{\echui}{\end{chui}}
\newcommand{\beq}{\begin{equation}}
\newcommand{\eeq}{\end{equation}}
\newcommand{\beqn}{\begin{eqnarray}}
\newcommand{\eeqn}{\end{eqnarray}}
\newcommand{\beqns}{\begin{eqnarray*}}
\newcommand{\eeqns}{\end{eqnarray*}}

\newcommand{\map}{\vph_n: \cX^n \to \cY^n}
\newcommand{\mapMtoY}{\vph_n: \cM_{M_n} \to \cY^n}
\newcommand{\mapXtoM}{\vph_n: \cX^n \to \cM_{M_n}}

\title{~\\~\\~\\~\\ \bf  Joint Source-Channel Coding Revisited: Information-Spectrum Approach
\thanks{This paper is an extended refinement of  a part of Chapter 3 in the book  Han \cite{Ha98}.}}
\author{~\\~\\~\\~\\Te Sun HAN\thanks{Te Sun Han was with  Graduate
School of Information Systems, University of Electro-Communications,
Chofugaoka 1-5-1, Chofu, Tokyo 182-8585, Japan. He is now visiting Department of Computer Science, Faculty of Science and Engineering, Waseda University, 
Room 902, Bld. 201 (Shinjuku Lamdax Building), Ohkubo 2-4-12, Shinjuku-ku, 
Tokyo 169-0072, Japan.
E-mail:\ han@is.uec.ac.jp, han@aoni.waseda.jp}}

\date{\today}

\begin{document}

\setcounter{page}{0}
\maketitle
\thispagestyle{empty}
\newpage

\pagenumbering{roman}

%
%

\pagenumbering{arabic}
\setcounter{page}{0}
\setcounter{equation}{0}

\vspace*{3cm}

{\bf Abstract:} \  Given a general source $\sV=\{V^n\}\noi$ with 
{\em countably infinite} source alphabet and a general channel 
$\sW=\{W^n\}\noi$ with arbitrary {\em abstract} channel input/channel output alphabets, we study the joint source-channel coding problem from the information-spectrum point of view.
First, we generalize Feinstein's lemma (direct part) and Verd\'u-Han's lemma (converse part) so as to be applicable  to the general joint source-channel coding problem. Based on these lemmas, 
we establish a sufficient condition as well as a necessary condition 
for the source $\sV$ to be reliably transmissible over the channel $\sW$
with asymptotically vanishing probability of error.
It is shown that our sufficient condition is equivalent to  the sufficient condition
derived by Vembu, Verd\'u and Steinberg \cite{vembu-verdu-stein}, whereas our necessary condition
is   shown  to be stronger  than or equivalent to the necessary condition derived by them. 
It turns out, as a direct consequence,  that ``{\em separation principle}" in a relevantly generalized sense  holds for a wide class of sources and channels, as was shown in a quite 
dfifferent manner by Vembu, Verd\'u
and Steinberg \cite{vembu-verdu-stein}. 
It  should also be remarked that a nice duality is found between our necessary and 
sufficient conditions, whereas we cannot fully enjoy 
such a duality between the necessary condition and the sufficient condition by Vembu, Verd\'u
and Steinberg \cite{vembu-verdu-stein}.
In addition, we demonstrate a sufficient condition as well as a necessary condition for the $\vep$-transmissibility ($0\le \vep <1$).
Finally, the separation theorem of the traditional standard form is shown to hold for the class of sources and channels that satisfy the semi-strong converse property.

\vspace{1cm}


\newpage

%
%
\section{Introduction}
\label{ss:2.intro}

Given a source $\sV=\{V^n\}\noi$ and a channel $\sW=\{W^n\}\noi$, 
 {\em joint source-channel coding } means
that the encoder maps the output from the source directly to the channel input
({\em one step encoding}), where the probability of decoding error is required to vanish as block-length $n$ tends to $\infty$. 
In usual situations, however, the  joint  source-channel coding  
can be decomposed into separate {\em source coding} and {\em channel coding} ({\em two step encoding}). This two step encoding does not cause 
any disadvantages from the standpoint of asymptotically vanishing error probabilities, provided that the so-called  {\em Separation Theorem} holds. 

Typically, the traditional separation theorem, which we call the separation theorem in the {\em narrow sense},  states that if the infimum $R_f(\sV)$ of all achievable fixed-length coding rates 
for the source $\sV$ is
smaller than the  capacity $C(\sW)$ for the channel $\sW$, then 
the source $\sV$ is reliably transmissible by two step encoding over the channel $\sW$; whereas if $R_f(\sV)$ is larger than $C(\sW)$ then the
reliable transmission is  impossible. 
While the former statement is always true for any general  source $\sV$ and any general channel $\sW$, the latter statement is {\em not} always true.
Then, a very natural question may be raised for  what class of sources and channels and in what sense the separation theorem holds in general.

Shannon \cite{shannon-1948} has first shown that the separation theorem holds for the class of stationary memoryless sources and channels. Since then, this theorem  has received extensive attention by a number of researchers who have attempted
to prove versions that apply to more and more general classes of sources and channels. Among others, for example, Dobrushin \cite{dobrushin}, Pinsker \cite{pinsker},  and Hu \cite{huguoding} have studied the separation theorem problem in the framework of information-stable sources and channels.

Recently, on the other hand, Vembu, Verd\'u and Steinberg 
\cite{vembu-verdu-stein} have put forth this problem in a much more general
information-spectrum context with  general source $\sV$ and  general channel $\sW$. From the viewpoint of information spectra, they have generalized  the notion of separation theorem and shown that,  usually 
 in many cases even with $R_f(\sV) > C(\sW)$,  it is possible to reliably transmit the output of the source $\sV$ over the channel $\sW$.
Furthermore, in terms of information spectra, they have established  a sufficient condition for the  transmissibility as well as a necessary condition.
It should be noticed here that, in this general joint source-channel coding situation, what indeed matters is not the validity problem of the traditional type of separation theorems but 
the derivation problem of necessary and/or sufficient conditions for 
the  transmissibility from the  information-spectrum point of view.
 
However, while their sufficient condition looks simple and significantly tight,
their necessary condition does not  look quite close to  tight. 

The present paper was mainly motivated by the reasonable question why the forms of these two conditions look rather very different
from one another. 
First, in Section 
\ref{ss: spectrum-lemmas}, the basic tools to answer this question are established, i.e., 
 two fundamental lemmas: a  generalization of Feinstein's lemma
\cite{feinstein}
and a generalization of Verd\'u-Han's lemma \cite{verdu-han-capacity}, which provide with the very basis for the key results to be stated in the subsequent sections. These lemmas are  of  {\em dualistic}  information-spectrum forms, which is  
in nice accordance with the general joint source-channel coding framework.
In Section \ref{ss: spectrum-ns-theorems}, given a general source $\sV$ and 
a general channel $\sW$,  we establish, in terms of information-spectra,
a sufficient condition ({\em Direct theorem}) for the  transmissibility as well as a necessary condition ({\em Converse theorem}).  The forms of these two conditions are very close from each other, and ``fairly" coincides with one another, provided that we dare disregard some relevant asymptotically 
vanishing term. 

Next, we equivalently rewrite these conditions in the forms useful to see 
relations to the separation theorem. As a consequence, it turns out that
a separation-theorem-like equivalent  of our sufficient condition just coincides with the sufficient condition given by Vembu, Verd\'u and Steinberg 
\cite{vembu-verdu-stein}, whereas a separation-theorem-like equivalent  of our necessary condition is  shown to be strictly stronger than 
or equivalent to the necessary condition given by them.
Here it is pleasing to observe  that a nice duality is found between our necessary and 
sufficient conditions, whereas we cannot fully enjoy such a duality between the necessary condition and the sufficient condition by Vembu, Verd\'u
and Steinberg \cite{vembu-verdu-stein}.

On the other hand, in Section \ref{ss:ep-achiev}, we demonstrate a sufficient condition as well as a necessary condition for the $\vep$-transmissibility,
which is the generalization of the sufficient condition as well as the necessary condition as was shown in Section \ref{ss: spectrum-ns-theorems}.
Finally, in Section \ref{ss:SEpThTra-1}, we restrict the class of sources and channels to those that satisfy the strong converse property (or, more generally, the semi-strong converse property) to show 
that the separation theorem in the traditional sense holds for this class. 
%
\section{Basic Notation and Definitions} \label{ss:basic-definition}

In this preliminary section, we prepare the basic notation and definitions
which will be used in the subsequent sections.

\subsection{General Sources}\label{sss:2.1}

Let us first give here the formal defintion of the  general source. A {\em general sources} is defined  as an infinite sequence
$\spr{V}=\{V^{n}=(V_{1}^{(n)},\cdots,V_{n}^{(n)})\}_{n=1}^{\infty}$ of
$n$-dimensional random variables $V^{n}$ where each component random
variable
$V_{i}^{(n)} \mbox{\enskip} (1 \leq i \leq n)$ takes values in a {\em
countably infinite}  set
$\cV$ that we call the {\em source alphabet}.
It should be noted here that each component of $V^{n}$ may change
depending on
block length $n$.
This implies that the sequence $\spr{V}$ is quite general in the sense that
it
may not satisfy even the consistency condition as usual processes,
where the consistency condition means that for any integers $m, n$ such that
$m < n$ it holds that $V^{(m)}_i \equiv V^{(n)}_i$ for all $ i = 1, 2,
\cdots, m.$
 The class of sources thus defined covers a very wide range of sources
including all nonstationary and/or nonergodic sources (cf. Han and Verd\'u \cite{han-verdu-approx}). 

\subsection{General Channels}\label{sss:2.2}

The formal definition of a general channel is as follows.
Let $\cX, \cY$ be arbitrary {\em abstract } ({\em not} necessarily countable)  sets, which we call 
the {\em input alphabet} and the {\em output alphabet}, respectively. 
A {\em general channel} is defined as an infinite sequence
$\sW =\{W^n:\cX^n\to\cY^n\}\noi$ of  $n$-dimensional 
probability transition matrices $W^n$, where $W^n(\ssy|\ssx) \ (\ssx \in \cX^n, 
\ssy \in \cY^n)$ denotes the conditonal probability of $\ssy$ given $\ssx$.\footnote{In the case where the output alphabet $\cY$ is {\em abstract}, $W^n(\ssy|\ssx)$ is understood to be the (conditional) probability measure element $W^n(d\ssy|\ssx)$ that is  measurable in $\ssx$.}  The class of channels thus defined covers a very wide range of channels including all nonstationary and/or nonergodic channels with arbitrary memory structures (cf. Han and Verd\'u \cite{han-verdu-approx}).

\bchui\label{chui:hajimeni-1}
{\rm 
A more reasonable definition of a general source is the following. 
Let $\{\cV_n\}\noi$ be any sequence of {\em arbitrary} source alphabets
$\cV_n$ (a  countabley infinite or abstract set) and let $V_n$ be any random variable taking values in $\cV_n$ $(n=1,2,\cdots)$.
Then, the sequence $\sV=\{V_n\}\noi$ of random variables $V_n$ is called
a {\em general source} (cf. Verd\'u and Han \cite{verdu-han-aep}). The above definition is a special case of this general source with $\cV_n = \cV^n$ $(n=1,2,\cdots)$. 

On the other hand, a more reasonable definition of the general channel is the following. Let $\{W_n:\cX_n\to\cY_n\}\noi$ be any sequence of {\em arbitrary} probability transition matrices, where $\cX_n, \cY_n$ are arbitrary abstract sets. Then, the sequence $\sW=\{W_n\}\noi$ of probability transition matrices $W_n$ is called
a {\em general channel} (cf. Han \cite{Ha98}). The above definition is a special case of this general channel with $\cX_n = \cX^n, \cY_n = \cY^n$ $(n=1,2,\cdots)$. 

 The  results in this paper
(Lemma \ref{hodai:source-channel_Direct}, Lemma \ref{hodai:source-channel_Converse}, Theorem \ref{teiri:3.17-1j},
Theorem \ref{teiri:3.17-2j}, 
Theorem \ref{teiri:3.17-3j},
Theorem \ref{teiri:3.17-4j}, Theorem \ref{teiri:3.17-1j-kruger},
Theorem \ref{teiri:3.17-2j-kruger} and Theorems \ref{teiri:Han_Verdu}
 $\sim$ \ref{teiri:3.17-5j_MW} ) 
continue to be valid  as well also in this more general setting with 
$\cV^n, V^n, \sV$ and $\cX^n, \cY^n, W^n, \sW$  replaced by 
$\cV_n$, $V_n, \sV$ and $\cX_n, \cY_n, W_n, \sW$, respectively.

In the sequel we use the convention that $P_Z(\cdot)$ denotes the probability distribution of a random variable $Z$, whereas $P_{Z|U}(\cdot|\cdot)$
denotes the conditional probability distribution of a random variable $Z$
given a random variable $U$.
\QED
}
\echui

\subsection{Joint Source-Channel Coding}\label{sss:2.3}

Let  $\spr{V}=\{V^{n}=(V_{1}^{(n)},\cdots,V_{n}^{(n)})\}_{n=1}^{\infty}$ be any general source, and let $\sW =\{W^n(\cdot|\cdot):\cX^n\to\cY^n\}\noi$ be any general channel. We consider  an {\em encoder}
$\varphi_n: \cV^n \to \cX^n$ and a {\em decoder}  $\psi_n: \cY^n \to \cV^n$, and put $X^n =\varphi_n(V^n)$. Then, denoting  by $Y^n$ the output from the channel $W^n$ due to the input $X^n$, we have
the obvious relation:

\beq\label{eq:Tora-1}
V^n\to X^n \to Y^n\quad (\mbox{a Markov chain}).
\eeq
The {\em error probability}  $\vep_n$  with  code 
$(\varphi_n, \psi_n)$  is defined by
\beqn\label{eq:Tora-2}
\vep_n &  \equiv & \Pr\left\{V^n \neq \psi_n(Y^n)\right\}\nonumber\\
& = &
 \sum_{\ssv \in \cV^n}P_{V^n}(\ssv)W^n(\cD^c(\ssv)|\varphi_n(\ssv)),
\label{eq:3.112-2j}
\eeqn
where  $\cD(\ssv) \equiv \{\ssy \in \cY^n | \psi_n(\ssy) =
\ssv\}$ $(\forall \ssv \in \cV^n)$ ($\cD(\ssv)$ is called the {\em decoding set } for $\ssv$) and ``$c$"  denotes the complement of a set. 
A pair $(\varphi_n, \psi_n)$ with error probability $\vep_n$   is simply called
 a joint source-channel  code $(n, \vep_n)$. 

We now define the {\em transmissibility} in terms of  joint source-channel codes 
$(n, \vep_n)$ as
\begin{teigi} \label{teigi:3.13-1j}
{\rm 
\mbox{}
\begin{eqnarray*}
\mbox{Source $\sV$ is transmissible over channel $\sW$}
& \defarrow & \mbox{There exists an $(n,\varepsilon_n)$ code }\\
& &\mbox{such that ${\displaystyle \lim_{n \rightarrow \infty}}\varepsilon_n = 0$}.
\end{eqnarray*}
}
\end{teigi}
With this definition of transmissibility,  in the following sections we shall establish  a sufficient condition as well as  a necessary condition for the transmissibility when we are given a geneal source $\sV$ and a general channel $\sW$. These two conditions are {\em very close} to each other and 
 could actually be seen  as giving ``{\em almost the same condition,}"  provided that  we dare disregard an asymptotically negligible term $\gamma_n \to 0$ appearing in those conditions (cf. Section \ref{ss: spectrum-ns-theorems}).
\bchui\label{chui:STAzealand1}
{\rm
The quantity $\varepsilon_n$ defined by (\ref{eq:3.112-2j}) 
is more specifically called the {\em average} error probability,
because it is averaged with respect to $P_{V^n}(\ssv)$ over all source outputs $\ssv \in \cV^n.$
On the other hand, we may define another kind of error probability by
\beq\label{eq:pamp1}
\varepsilon_n \equiv \sup_{\ssv : P_{V^n}(\ssv)>0}W^n(\cD^c(\ssv)|\varphi_n(\ssv)),
\eeq
which we call the {\em maximum} error probability. 
It is evident that the transmissibility in the maximum sense implies the transmissibility in the 
average sense. However, the inverse is not necessarily true. To see this, it suffices 
to consider the following simple example.
Let the source, channel input, channel output alphabets be
$\cV_n = \{0, 1, 2\},$ $\cX_n = \{1,2\}, $ $\cY_n = \{1,2\}$, respectively;
and the (deterministic) channel $W_n: \cX_n \to \cY_n$ be defined by 
$W_n(j|i)=1$ for $i = j, W_n(1|0) =1$.
Moreover, let the source $V_n$ have  probability distribution
$P_{V_n}(0) = \alpha_n, $  $P_{V_n}(1) = P_{V_n}(2) =$ $ \frac{1-\alpha_n}{2}$ 
($\alpha_n \to 0$ as $n\to \infty $).
One of the best choices of  possible pairs of encoder-decoder $(\varphi_n: \cV_n\to \cX_n, 
\ \psi_n: \cY_n\to\cV_n)$, either in the average sense or in the maximum sense,  is such
that $\varphi_n(i)=i$ for $i=1,2; \varphi_n(0)=1;$
 $\psi_n(i)=i$ for $i=1,2$.
 Then, the average error probability is  $\varepsilon_n^{\mbox{{\scriptsize a}}}= \alpha_n
 \to 0,$
 while the maximum error probability is 
 $\varepsilon_n^{\mbox{{\scriptsize m}}} = 1.$
Thus, in this case, the source $V_n$ is transmissible in the average sense over the channel $W_n$, while
it is {\em not} transmissible in the maximum sense.

Hereafter, the probability  $\varepsilon_n$ is understood to denote the ``average" error probability, unless otherwise stated.
\QED
}  
\echui
 \section{Fundamental Lemmas\label{ss: spectrum-lemmas}}
 In this section, we prepare two fundamental lemmas that are needed 
 in the next section in order to establish the main theorems ({\em Direct part} and {\em Converse part}).
\begin{hodai}[Generalization of Feinstein's lemma]
 \label{hodai:source-channel_Direct}
 {\rm  Given a general  
   source 
  $\sV=\{V^n\}\noi$ and a general channel 
 $\sW=\{W^n\}\noi$, 
let $X^n$ be any input random variable taking values in $\cX^n$  and $Y^n$ be the channel output via $W^n$ due to  the channel input $X^n$, where $V^n\to X^n \to Y^n$. Then, for every 
$n = 1, 2, \cdots$, there exists an $(n, \vep_n)$ code such that 
\beq\label{eq:3.112-3j}
\vep_n \le \Pr\left\{ \MID \le \nth \log\frac{1}{P_{V^n}(V^n)}
 + \gamma \right\} + e^{-n \gamma},
\eeq
where\footnote{In the case where the input and output alphabets
$\cX, \cY$ are {\em abstract} ({\em not} necessarily countable),
 $\frac{W^n(Y^n|X^n)}{P_{Y^n}(Y^n)}$ in (\ref{eq:3.112-3j}) is understood to be $g(Y^n|X^n)$, where $g(\ssy | \ssx) \equiv
 \frac{W^n(d\ssy|\ssx)}{P_{Y^n}(d\ssy)}$  
 $=\frac{W^n(d\ssy|\ssx)P_{X^n}(d\ssx)}{P_{Y^n}(d\ssy)P_{X^n}(d\ssx)}$  $=\frac{P_{X^nY^n}(d\ssx,d\ssy)}{P_{X^n}(d\ssx)P_{Y^n}(d\ssy)}$ is the Radon-Nikodym derivative
that is measurable in  $(\ssx,\ssy)$.} $\gamma>0$ is an arbitrary positive number.
}
\end{hodai}
\bchui\label{chui:3.13-1j}
{\rm
In a special case where the source $\sV =\{V^n\}\noi$ is uniformly distributed on the massage set 
$\cM_n = \left\{1,2,\cdots, M_n\right\}$, it follows that 
\[
\nth\log\frac{1}{P_{V^n}(V^n)} = \nth \log M_n,
\]
which implies that the entropy spectrum\footnote{The probablity distribution of $\nth \log\frac{1}{P_{V^n}(V^n)}$ is called the {\em entropy spectrum} of the source $\sV=\{V^n\}\noi$, whereas the probability distribution of $\MID$ is called the {\em mutual information spectrum} of the channel $\sW=\{W^n\}\noi$ given the input $\sX=\{X^n\}\noi$ (cf. Han and Verd\'u \cite{han-verdu-approx}).} of the source $\sV =\{V^n\}\noi$ is exactly  one point spectrum concentrated on $\nth \log M_n$. Therefore, in this special case, Lemma 
\ref{hodai:source-channel_Direct} reduecs to  
Feinstein's lemma \cite{feinstein}.
}
\chuiowari
\echui

\noindent
{\em Proof of Lemma\ref{hodai:source-channel_Direct}}:

\bigskip

For each $\ssv \in \cV^n$, generate $\ssx(\ssv) \in \cX^n$ at random according to the conditional distribution $P_{X^n|V^n}(\cdot|\ssv)$ and let 
$\ssx(\ssv)$ be the codeword for  $\ssv$. In other words, we define the encoder $\varphi_n:\cV^n\to \cX^n$ as $\varphi_n(\ssv) = \ssx(\ssv)$,
where $\{\ssx(\ssv)\  | \ \forall \ssv \in \cV^n\}$ are all independently generated. We define the decoder $\psi_n: \cY^n \to\cV^n$ as follows: Set
\beqn
S_n
  & = & \left\{(\ssv,\ssx, \ssy)\in \cZ^n 
\left| \nth\log\frac{W^n(\ssy | \ssx)}{\pyn(\ssy)} 
> \nth \log\frac{1}{P_{V^n}(\ssv)} + \gamma \right.\right\},\nonumber\\
& &  \label{eq:3.112-4j}\\
S_n(\ssv) & =  & \left\{(\ssx, \ssy)\in \cX^n \times \cY^n 
\left| (\ssv,\ssx, \ssy)\in S_n \right.\right\},\label{eq:3.112-5j}
\eeqn
where for simplicity we have put
$\cZ^n \equiv  \cV^n \times \cX^n \times \cY^n$.
Suppose that the decoder 
$\psi_n$ received a channel output $\ssy \in \cY^n$. If there exists one and only one 
$\ssv \in \cV^n$ such that
$(\ssx(\ssv), \ssy) \in S_n(\ssv)$, define the decoder as 
$ \psi_n(\ssy) = \ssv $; otherwise,  let the output of the decoder  $\psi_n(\ssy) \in \cV^n$ be arbitrary. Then, the probability  $\overline{\vep}_n$ of error for this pair
$(\varphi_n, \psi_n)$  (averaged over all the realizatioins of the random code) is given by 
\beq\label{eq:3.112-6j}
\overline{\vep}_n =
 \sum_{\ssv \in \cV^n}P_{V^n}(\ssv)\overline{\vep}_n(\ssv),
\eeq
where $\overline{\vep}_n(\ssv)$ is the probability of error (averaged over all the realizatioins of the random code)  when   $\ssv \in \cV^n$ is
the source output. 
We can evaluate 
$\overline{\vep}_n(\ssv)$ as
\beqn
\overline{\vep}_n(\ssv) &  \le & \Pr\left\{(\ssx(\ssv), Y^n) \notin S_n(\ssv)\right\}
\nonumber\\
&  & + \Pr\left\{\bigcup_{\ssv': \ssv' \neq \ssv}
\left\{(\ssx(\ssv'), Y^n) \in S_n(\ssv')\right\}\right\}\nonumber\\
& \le &  \Pr\left\{(\ssx(\ssv), Y^n) \notin S_n(\ssv)\right\}
\nonumber\\ 
& & + \sum_{\ssv': \ssv' \neq \ssv}
\Pr\left\{(\ssx(\ssv'), Y^n) \in S_n(\ssv')\right\},
\label{eq:3.112-7j}
\eeqn
where $Y^n$ is the channel output via $W^n $ due to  the channel input $\ssx(\ssv)$.
The first term on the right-hand side of (\ref{eq:3.112-7j}) is written as
\beqns
A_n(\ssv) & \equiv &  \Pr\left\{(\ssx(\ssv), Y^n) \notin S_n(\ssv)\right\}\\
&= & \sum_{(\ssx, \ssy) \notin S_n(\ssv)}
P_{X^nY^n|V^n}(\ssx,\ssy|\ssv).
\eeqns
Hence,
\beqn
\sum_{\ssv \in \cV^n}P_{V^n}(\ssv)A_n(\ssv)
& = & \sum_{\ssv \in \cV^n}P_{V^n}(\ssv)
\sum_{(\ssx, \ssy) \notin S_n(\ssv)}
P_{X^nY^n|V^n}(\ssx,\ssy|\ssv)\nonumber\\
& = & \sum_{(\ssv, \ssx, \ssy) \notin S_n}
P_{V^nX^nY^n}(\ssv, \ssx,\ssy)\nonumber\\
& = & \Pr\left\{V^nX^nY^n \notin S_n\right\}.\label{eq:3.112-8j}
\eeqn
On the other hand, noting that $\ssx(\ssv'), \ssx(\ssv)$ $(\ssv' \neq \ssv)$ are independent and hence $\ssx(\ssv')$, $Y^n$ are also independent, 
 the second term on the right-hand side of (\ref{eq:3.112-7j}) is evaluated as 
\beqns
B_n(\ssv) & \equiv &\sum_{\ssv': \ssv' \neq\ssv}
\Pr\left\{(\ssx(\ssv'), Y^n) \in S_n(\ssv')\right\}\\
& = & \sum_{\ssv': \ssv' \neq\ssv}
\sum_{(\ssx, \ssy) \in S_n(\ssv')}P_{Y^n|V^n}(\ssy|\ssv)
P_{X^n|V^n}(\ssx|\ssv')\\
& \le & \sum_{\ssv' \in \cV^n}\sum_{(\ssx, \ssy) \in S_n(\ssv')}
P_{Y^n|V^n}(\ssy|\ssv)P_{X^n|V^n}(\ssx|\ssv').
\eeqns
Hence,
\beqn
\lefteqn{\sum_{\ssv \in \cV^n}P_{V^n}(\ssv)B_n(\ssv)}\nonumber\\
& \le & \sum_{\ssv \in \cV^n}\sum_{\ssv' \in \cV^n}
\sum_{(\ssx, \ssy) \in S_n(\ssv')}
P_{V^n}(\ssv)P_{Y^n|V^n}(\ssy|\ssv)P_{X^n|V^n}(\ssx|\ssv')\nonumber\\
& = & 
\sum_{\ssv' \in \cV^n}
\sum_{(\ssx, \ssy) \in S_n(\ssv')}
P_{Y^n}(\ssy)P_{X^n|V^n}(\ssx|\ssv').
\label{eq:3.112-9j}
\eeqn
On the other hand, in view of 
(\ref{eq:3.112-4j}), (\ref{eq:3.112-5j}), 
$(\ssx, \ssy) \in S_n(\ssv')$ implies
\[
P_{Y^n}(\ssy) \le 
P_{V^n}(\ssv')W^n(\ssy|\ssx)e^{-n\gamma}.
\]
Therefore, 
(\ref{eq:3.112-9j}) is further transformed to
\beqn
\lefteqn{\sum_{\ssv \in \cV^n}P_{V^n}(\ssv)B_n(\ssv)}\nonumber\\
& \le & e^{-n\gamma}\sum_{\ssv' \in \cV^n}
\sum_{(\ssx, \ssy) \in S_n(\ssv')}
P_{V^n}(\ssv')P_{X^n|V^n}(\ssx|\ssv')W^n(\ssy|\ssx)\nonumber\\
& \le &  e^{-n\gamma}\sum_{(\ssv', \ssx, \ssy) \in \cZ^n}
P_{V^n}(\ssv')P_{X^n|V^n}(\ssx|\ssv')W^n(\ssy|\ssx)\nonumber\\
& = & e^{-n\gamma}.\label{eq:3.112-10j}
\eeqn
Then, from (\ref{eq:3.112-6j}),
(\ref{eq:3.112-8j}) and (\ref{eq:3.112-10j}) it follows that
\beqns
\overline{\vep}_n & = &
 \sum_{\ssv \in \cV^n}P_{V^n}(\ssv)\overline{\vep}_n(\ssv)\\
& \le & \sum_{\ssv \in \cV^n}P_{V^n}(\ssv)A_n(\ssv) +
\sum_{\ssv \in \cV^n}P_{V^n}(\ssv)B_n(\ssv)\\
& \le & \Pr\left\{V^nX^nY^n \notin S_n\right\} + e^{-n\gamma}.
\eeqns
Thus, there must exist a deterministic 
$(n,\vep_n)$ code such that
\[
\vep_n \le \Pr\left\{V^nX^nY^n \notin S_n\right\} + e^{-n\gamma},
\]
thereby proving Lemma \ref{hodai:source-channel_Direct}. \QED

\begin{hodai}[Generalization of Verd\'u-Han's lemma]
\label{hodai:source-channel_Converse}
{\rm
Let $\sV=\{V^n\}\noi$ and $\sW=\{W^n\}\noi$ be a general source and a general channel, respectively, and  let  $\varphi_n: \cV^n\to\cX^n$ be the encoder  of an $(n, \vep_n)$ code for $(V^n,$ $ W^n)$. Put $X^n = \varphi_n(V^n)$
and let $Y^n$ be the channel output via $W^n$ due to the channel input 
$X^n$, where $V^n\to X^n \to Y^n$. Then, for every $n =1, 2, \cdots$, it holds that
\beq\label{eq:3.112-13j}
\vep_n \ge \Pr\left\{\MID \le 
\nth\log\frac{1}{P_{V^n}(V^n)} -\gamma \right\} - e^{-n\gamma},
\eeq
where $\gamma > 0$ is an arbitrary positive number.
}
\end{hodai}

\begin{chui}
{\rm
In a special case where the source $\sV =\{V^n\}\noi$ is uniformly distributed on the massage set 
$\cM_n = \left\{1,2,\cdots, M_n\right\}$, it follows that 
\[
\nth\log\frac{1}{P_{V^n}(V^n)} = \nth \log M_n,
\]
which implies that the entropy spectrum of the source $\sV =\{V^n\}\noi$ is exactly  one point spectrum concentrated on $\nth \log M_n$. Therefore, in this special case, Lemma 
\ref{hodai:source-channel_Converse} reduecs to  
Verd\'u-Han's lemma \cite{verdu-han-capacity}.
}
\QED
\end{chui}

\noindent
{\em Proof of Lemma\ref{hodai:source-channel_Converse}}

\bigskip

Define
\beq\label{eq:3.112-14j}
L_n = \left\{(\ssv,\ssx, \ssy)\in \cZ^n\left | \nth \log\frac{W^n(\ssy|\ssx)}
{P_{Y^n}(\ssy)} \le 
\nth\log\frac{1}{P_{V^n}(\ssv)} -\gamma\right.\right\},
\eeq    
and, for each $\ssv \in \cV^n$  set
\[
\cD(\ssv) = \left\{\ssy \in \cY^n | \psi_n(\ssy) = \ssv\right\},
\]
that is,  $\cD(\ssv)$ is  the decoding set for $\ssv$.
Moreover, for each $(\ssv, \ssx) \in $
$\cV^n \times \cX^n$, set
\beq\label{eq:3.112-15j}
\cB(\ssv,\ssx) = \left\{\ssy \in \cY^n | (\ssv, \ssx, \ssy) \in L_n\right\}.
\eeq
Then, noting the Markov chain property (\ref{eq:Tora-1}), we have
\beqn
\lefteqn{\Pr\left\{V^nX^nY^n \in L_n\right\}}\nonumber\\
& = & \sum_{(\ssv, \ssx, \ssy) \in L_n}P_{V^nX^nY^n}(\ssv, \ssx, \ssy)
\nonumber\\
& = & \sum_{(\ssv, \ssx) \in \cV^n \times \cX^n}
P_{V^nX^n}(\ssv, \ssx)
W^n(\cB(\ssv, \ssx) | \ssx)\nonumber\\
& = & \sum_{(\ssv, \ssx) \in \cV^n \times \cX^n}
P_{V^nX^n}(\ssv, \ssx)
W^n(\cB(\ssv, \ssx)\cap \cD^c(\ssv) | \ssx)\nonumber\\
& & +  \sum_{(\ssv, \ssx) \in \cV^n \times \cX^n}
P_{V^nX^n}(\ssv, \ssx)W^n(\cB(\ssv, \ssx)\cap \cD(\ssv) | \ssx)
\nonumber\\
& \le &\sum_{(\ssv, \ssx) \in \cV^n \times \cX^n}
P_{V^nX^n}(\ssv, \ssx)
W^n( \cD^c(\ssv) | \ssx)\nonumber\\
& & +  \sum_{(\ssv, \ssx) \in \cV^n \times \cX^n}
P_{V^nX^n}(\ssv, \ssx)W^n(\cB(\ssv, \ssx)\cap \cD(\ssv) | \ssx)
\nonumber\\
& = & \vep_n +  \sum_{(\ssv, \ssx) \in \cV^n \times \cX^n}
P_{V^nX^n}(\ssv, \ssx)W^n(\cB(\ssv, \ssx)\cap \cD(\ssv) | \ssx)
\nonumber\\
& = & \vep_n + 
 \sum_{(\ssv, \ssx) \in \cV^n \times \cX^n}
P_{V^nX^n}(\ssv, \ssx)\sum_{\ssy \in \cB(\ssv, \ssx)\cap \cD(\ssv)}
W^n(\ssy|\ssx),
\label{eq:3.112-16j}
\eeqn
where we have used the relation:
\[
\vep_n = \sum_{(\ssv, \ssx) \in \cV^n \times \cX^n}
P_{V^nX^n}(\ssv, \ssx)W^n(\cD^c(\ssv)|\ssx).
\]
Now, it follows from (\ref{eq:3.112-14j}) and  (\ref{eq:3.112-15j})  that 
$\ssy \in \cB(\ssv,\ssx)$ implies
\[
W^n(\ssy|\ssx) \le \frac{e^{-n\gamma}\pyn(\ssy)}{P_{V^n}(\ssv)},
\]
which is substituted into the right-hand side of (\ref{eq:3.112-16j}) to yield
\beqns
\lefteqn{\Pr\left\{V^nX^nY^n \in L_n\right\}}\\
& \le & \vep_n +
e^{-n\gamma}\sum_{(\ssv, \ssx) \in \cV^n \times \cX^n}
P_{X^n|V^n}(\ssx|\ssv)
\sum_{\ssy \in \cB(\ssv, \ssx)\cap \cD(\ssv)}\pyn(\ssy)\\
& \le & \vep_n +
e^{-n\gamma}\sum_{(\ssv, \ssx) \in \cV^n \times \cX^n}
P_{X^n|V^n}(\ssx|\ssv)\pyn(\cD(\ssv))\\
& = & \vep_n +
e^{-n\gamma}\sum_{\ssv \in \cV^n}
\pyn(\cD(\ssv))\nonumber\\
& = & \vep_n + e^{-n\gamma},
\eeqns
thereby proving the claim of the lemma.
\QED
\section{Theorems on Transmissibility}\label{ss: spectrum-ns-theorems}

In this section we give both of a sufficient condition and a necessary condition
for the transmissibility with a given general souce $\sV=\{V^n\}\noi$ and a
given general channel $\sW=\{W^n\}\noi$.

First, Lemma \ref{hodai:source-channel_Direct} immediately leads us to the following direct theorem:

\begin{teiri}[Direct theorem]\label{teiri:3.17-1j}
{\rm
Let $\sV =\{V^n\}\noi$, $\sW =\{W^n\}\noi$ be a general source and a general channel, respectively. If there exist {\em some} channel input 
$\sX =\{X^n\}\noi$
and  {\em some} sequence $\{\gamma_n\}\noi$
satisfying 
\beq\label{eq:3.112-11j}
\gamma_n > 0, \ \gamma_n \to 0\ \mbox{and}\ n\gamma_n \to \infty \quad (n\to\infty)
\eeq
for which it holds that
\beq\label{eq:3.112-12j}
\lmn \Pr\left\{\MID \le 
\nth\log\frac{1}{P_{V^n}(V^n)} +\gamma_n \right\}= 0,
\eeq
then  the source $\sV =\{V^n\}\noi$ is transmissible over the channel 
$\sW =\{W^n\}\noi$, where 
$Y^n$ is the channel output via $W^n $ due to  the channel input $X^n$
and $V^n\to X^n\to Y^n$.
}
\end{teiri}

\noindent
{\em Proof:}

Since in Lemma \ref{hodai:source-channel_Direct} we can choose the constant $\gamma > 0$ so as to depend on $n$,
let us take, instead of $\gamma$, an arbitrary $\{\gamma_n\}\noi$ satisfying 
 condition 
(\ref{eq:3.112-11j}). Then, the second term on the right-hand side of 
(\ref{eq:3.112-3j}) vanishes as $n$ tends to $\infty$, and hence it follows from (\ref{eq:3.112-12j}) that the right-hand side of 
(\ref{eq:3.112-3j}) vanishes as $n$ tends to $\infty$.
Therefore, the $(n, \vep_n)$ code as specified in 
Lemma \ref{hodai:source-channel_Direct} satisfies 
${\dss \lmn \vep_n = 0}$. \QED

Next, Lemma \ref{hodai:source-channel_Converse} immediately leads us to the following converse theorem:

\begin{teiri}[Converse theorem]\label{teiri:3.17-2j}
{\rm
Suppose that  a general source $\sV =\{V^n\}\noi$ is transmissible over a general channel $\sW =\{W^n\}\noi$. 
Let the channel input be $\sX =\{X^n \equiv \varphi_n(V^n)\}\noi$
where $\varphi_n: \cV^n\to \cX^n$
is the channel encoder.
Then, for {\em any} sequence $\{\gamma_n \}\noi$
satisfying  condition (\ref{eq:3.112-11j}), it holds that
\beq\label{eq:3.112-17j}
\lmn \Pr\left\{\MID \le 
\nth\log\frac{1}{P_{V^n}(V^n)} -\gamma_n \right\}=0,
\eeq
where $Y^n$ is the channel output via $W^n $ due to  the channel input $X^n$ and $V^n\to X^n\to Y^n$.}
\end{teiri}

\noindent
{\em Proof:}

If $\sV$ is transmissible over $\sW$, then, by Definition \ref{teigi:3.13-1j} there exists an $(n, \vep_n)$ code such that
${\dss \lmn\vep_n = 0}$.
 Hence,  the claim of the theorem immediately follows from (\ref{eq:3.112-13j}) in Lemma 
\ref{hodai:source-channel_Converse} with $\gamma_n$ instead of 
$\gamma$.
\QED

\bchui\label{chui:3.13-3j}
{\rm
Comparing (\ref{eq:3.112-17j}) in Theorem \ref{teiri:3.17-2j} with 
(\ref{eq:3.112-12j}) in Theorem \ref{teiri:3.17-1j},
we observe that the only difference is that the sign of $\gamma_n$ is changed 
from 
$+$ to $-$.  Since $\gamma_n$ vanishes as $n$ tends to $\infty$, this difference is asymptotically negligible. 
}
\chuiowari
\echui

Now, let us think of the implication of  conditions (\ref{eq:3.112-12j})
and (\ref{eq:3.112-17j}).  First, let us think of (\ref{eq:3.112-12j}).
Putting
\[
A_n = \MID , \quad B_n = \nth\log\frac{1}{P_{V^n}(V^n)}
\]
for simplicity, (\ref{eq:3.112-12j}) is written as 
\beq\label{eq:3.112-18j}
\alpha_n \equiv \Pr\left\{A_n \le B_n +\gamma_n\right\}
 \to 0 \quad (n\to\infty),
\eeq
which can be  transformed to
\beqns
\lefteqn{\Pr\left\{A_n \le B_n +\gamma_n\right\}}\\
& = & \sum_{u}\Pr\left\{B_n = u\right\}\Pr\left\{A_n \le B_n +\gamma_n |B_n = u\right\}\\
& = &  \sum_{u}\Pr\left\{B_n = u\right\}\Pr\left\{A_n \le u +\gamma_n |B_n = u\right\}.
\eeqns
Set
\beq\label{eq:3.112-19j}
T_n = \left\{ u \mid \Pr\left\{A_n \le u +\gamma_n |B_n = u\right\}
 \le \sqrt{\alpha_n}\right\},
\eeq
then by virtue of (\ref{eq:3.112-18j}) and Markov inequality, we have 
\beq\label{eq:3.112-20j}
\Pr\left\{B_n \in T_n\right\} \ge 1 - \sqrt{\alpha_n}.
\eeq
Let us now define the upper cumulative probabilities for $A_n, B_n$ by
\[
P_n(t) = \Pr\left\{A_n \ge t\right\}, \quad Q_n(t) = \Pr\left\{B_n \ge t\right\},
\]
then it follows that 
\beqn
P_nitj& = & \sum_{u}\Pr\left\{B_n = u\right\}\Pr\left\{A_n \ge t |B_n = u\right\}
\nonumber\\
& \ge &  \sum_{\stackrel{u \in T_n:}{u \ge t - \gamma_n}}
\Pr\left\{B_n = u\right\}\Pr\left\{A_n \ge t |B_n = u\right\}\nonumber\\
& \ge &  \sum_{\stackrel{u \in T_n:}{u \ge t - \gamma_n}}
\Pr\left\{B_n = u\right\}\Pr\left\{A_n \ge u + \gamma_n |B_n = u\right\}.
\label{eq:3.112-21j}
\eeqn
On the other hand, by means of (\ref{eq:3.112-19j}), $u \in T_n$ implies 
that
\[
\Pr\left\{A_n \ge u + \gamma_n | B_n = u\right\} \ge 1 - \sqrt{\alpha_n}.
\]
Theore, by (\ref{eq:3.112-20j}), (\ref{eq:3.112-21j}) it is concluded that
\beqns
P_n(t) & \ge & (1 - \sqrt{\alpha_n})
 \sum_{\stackrel{u \in T_n:}{u \ge t - \gamma_n}}\Pr\left\{B_n = u\right\}\\
& \ge & (1 - \sqrt{\alpha_n})(Q_n(t - \gamma_n) - \Pr\left\{B_n \notin T_n\right\})\\
& \ge & (1 - \sqrt{\alpha_n})(Q_n(t - \gamma_n) - \sqrt{\alpha_n})\\
& \ge & Q_n(t - \gamma_n) - 2\sqrt{\alpha_n}.
\eeqns
That is, 
\[
P_n(t) \ge Q_n(t - \gamma_n) - 2\sqrt{\alpha_n}.
\]
This means
 that, for all $t$, the upper cumulative probability $P_n(t)$ of $A_n$ is larger than or equal to the upper cumulative probability $Q_n(t-\gamma_n)$ of 
$B_n$, except for the asymptotically vanishing difference $2\sqrt{\alpha_n}$. 
This in turn implies that, as a whole, the mutual information spectrum of the channel is shifted to the right in comparison with the entropy spectrum of the source.
 With  $-\gamma_n$ instead of $\gamma_n$, the same implication follows also from 
(\ref{eq:3.112-17j}). 
It is such an allocation relation between
the mutual information spectrum and the entropy spectrum that enables us to make an transmissible joint source-channel coding. 

\medskip

However, it is not easy in general to check whether  conditions 
(\ref{eq:3.112-12j}), (\ref{eq:3.112-17j}) in these forms are satisfied or not. Therefore, we consider to equivalently rewrite  conditions (\ref{eq:3.112-12j}), 
(\ref{eq:3.112-17j}) into alternative information-spectrum  forms  hopefully easier to depict an intuitive picture. This can actually be done by re-choosing the input and output variables
$X^n, Y^n$ as below. These forms are useful in order to see the relation 
of  conditions (\ref{eq:3.112-12j}), (\ref{eq:3.112-17j}) with the so-called 
{\em separation theorem}.

First, we show another information-spectrum form equivalent to the sufficient condition 
(\ref{eq:3.112-12j}) in Theorem \ref{teiri:3.17-1j}.

\begin{teiri}[Equivalence of sufficient conditions]\label{teiri:3.17-3j}
{\rm
The following two conditions are equivalent:
}

\medskip

{\rm
1) For {\em  some} channel input $\sX =\{X^n\}\noi$ 
 and 
  {\em  some} sequence $\{\gamma_n \}\noi$ satisfying  condition (\ref{eq:3.112-11j}), it holds that
\beq\label{eq:3.112-22j}
\lmn \Pr\left\{\MID \le 
\nth\log\frac{1}{P_{V^n}(V^n)} + \gamma_n \right\}=0,
\eeq
where $Y^n$ is the channel output via $W^n $ due to  the channel input $X^n$  and $V^n\to X^n\to Y^n$.
}

\medskip

{\rm
2) ({\bf Strict domination:}
 Vembu, Verd\'u and Steinberg \cite{vembu-verdu-stein})
 For {\em  some} channel input $\sX =\{X^n\}\noi$, 
{\em  some} sequence $\{c_n\}\noi$ and  {\em  some} sequence 
 $\{\gamma_n \}\noi$ satisfying  condition (\ref{eq:3.112-11j}), it holds that
\beqn
\lefteqn{\lmn \left(
\Pr\left\{\nth\log\frac{1}{P_{V^n}(V^n)} 
\ge c_n \right\}\right.}\nonumber\\
& & \quad\quad\quad  +  
\left.\Pr\left\{\MID \le c_n + \gamma_n\right\}\right) 
 = 0, \label{eq:3.112-23j}
\eeqn
where $Y^n$ is the channel output via $W^n $ due to  the channel input $X^n$.
}
\\
\end{teiri}

\bchui[separation in general]\label{chui:3.13-5j}
{\rm@
The sufficient condition 2) in Theorem \ref{teiri:3.17-3j} means that the entropy spectrum of the source  and the mutual information spectrum of the channel are asymptotically completely split with a vacant boundary  of  asymptotically vanishing width $\gamma_n$, and the former is placed to the left of the latter, where these two spectra may oscillate ``synchronously" with $n$. In the case where such a separation condition 2)  is satisfied, we can split  reliable joint source-channel coding in two steps as follows ({\em separation} of source coding and channel coding): We first encode the source output $V^n$ at the fixed-length coding rate
$ c_n = \nth \log M_n$ ($M_n$ is the size of the message set $\cM_n$), and then encode the output of the source encoder into the channel. The error probabilty $\vep_n$ for this two step coding is upper bounded by the sum of the
error probability of the 
fixed-length source coding (cf. Vembu, Verd\'u and  Steinberg \cite{vembu-verdu-stein};  Han \cite[Lemma 1.3.1]{Ha98}):
\[
\Pr\left\{\nth\log\frac{1}{P_{V^n}(V^n)} \ge c_n\right\}
\]
and the ``maximum" error probability of the channel coding (cf. Feinstein \cite{feinstein}, Ash 
\cite{ashtein}, Han \cite[Lemma 3.4.1]{Ha98}):
\[
\Pr\left\{\MID \le c_n + \gamma_n \right\} + e^{-n\gamma_n}.
\]
It then follows from (\ref{eq:3.112-23j}) that both of these two error probabilities vanish as $n$ tends to $\infty$, where it should be noted that 
$e^{-n\gamma_n}\to 0$ as $n\to\infty$. Thus, we have 
${\dss \lmn\vep_n = 0}$ to conclude that the source $\sV=\{V^n\}\noi$ is transmissible over the channel 
$\sW=\{W^n\}\noi$. This can be regarded as providing  another proof of Theorem 
\ref{teiri:3.17-1j}.
}
\chuiowari
\echui

\noindent
{\em Proof of Theorem \ref{teiri:3.17-3j}:}

\noindent
2) $\Rightarrow$ 1):
For any joint probability distribution $P_{V^nX^n}$ for $V^n$ and $X^n$, we have
\beqns
\lefteqn{\Pr\left\{\MID \le 
\nth\log\frac{1}{P_{V^n}(V^n)} + \gamma_n \right\}}\\
& \le & \Pr\left\{\nth\log\frac{1}{P_{V^n}(V^n)} 
\ge c_n \right\}\\
& & \quad +  
\Pr\left\{\MID \le c_n + \gamma_n\right\},
\eeqns
which together with (\ref{eq:3.112-23j}) implies
(\ref{eq:3.112-22j}). \\

\noindent
1) $\Rightarrow$ 2)F
Supposing that condition 1) holds, put
\beq\label{eq:3.112-24j}
\alpha_n \equiv 
\Pr\left\{\MID \le 
\nth\log\frac{1}{P_{V^n}(V^n)} + \gamma_n \right\},
\eeq
and moreover, with $ \gamma_n^{\prime} = \frac{\gamma_n}{4},$
$\delta_n = \max (\sqrt{\alpha_n}, e^{-n\gamma_n^{\prime}})$, define

\beq\label{eq:3.112-25j}
d_n = \sup \left\{R \left| \Pr\left\{\nth\log\frac{1}{P_{V^n}(V^n)} 
\ge R\right.\right\} > \delta_n\right\} - \gamma_n^{\prime}.
\eeq
Furthermore, define 
\beqn
S_n &  = & \left\{\ssv \in \cV^n \left| \nth\log\frac{1}{P_{V^n}(\ssv)} 
\ge d_n\right.\right\},\label{eq:3.112-26j}\\
\lambda_n^{(1)} &  =& \Pr\left\{V^n \in S_n\right\}, \quad 
\lambda_n^{(2)} = \Pr\left\{V^n \notin S_n\right\},\label{eq:3.112-26-1j}
\eeqn
then the joint probability distribution $P_{V^nX^nY^n}$ can be written as a mixture:
\beqn
\lefteqn{P_{V^nX^nY^n}(\ssv, \ssx, \ssy)}\nonumber\\
& = &
 \lambda_n^{(1)}P_{\tilde{V}^n\tilde{X}^n\tilde{Y}^n}(\ssv, \ssx, \ssy)
 + \lambda_n^{(2)}P_{\overline{V}^n\overline{X}^n\overline{Y}^n}
(\ssv, \ssx, \ssy),\label{eq:3.112-27j}
\eeqn
where $P_{\tilde{V}^n\tilde{X}^n\tilde{Y}^n},$
$P_{\overline{V}^n\overline{X}^n\overline{Y}^n}$ are the conditional probability distributions of $V^nX^nY^n$ conditioned on 
$V^n \in S_n,$ $V^n \notin S_n$, respectively.
We notice here that the Markov chain property $V^n\to X^n\to Y^n$ implies
$P_{\tilde{Y}^n|\tilde{X}^n} = $ $P_{\overline{Y}^n|\overline{X}^n}$ 
$ = W^n$ and the Markov chain properties
\[
\tilde{V}^n\to \tilde{X}^n\to \tilde{Y}^n,\quad 
\overline{V}^n\to \overline{X}^n\to \overline{Y}^n.
\]
We now rewrite 
(\ref{eq:3.112-24j}) as
\beqn
\alpha_n & = & 
\lambda_n^{(1)}
\Pr\left\{\nth\log\frac{W^n(\tilde{Y}^n| \tilde{X}^n)}{P_{Y^n}(\tilde{Y}^n)} \le 
\nth\log\frac{1}{P_{V^n}(\tilde{V}^n)} + \gamma_n \right\}\nonumber\\
& & +
\lambda_n^{(2)}\Pr\left\{\nth\log\frac{W^n(\overline{Y}^n| \overline{X}^n)}{P_{Y^n}(\overline{Y}^n)} \le 
\nth\log\frac{1}{P_{V^n}(\overline{V}^n)} + \gamma_n \right\}.\nonumber\\
& & 
\label{eq:3.112-28j}
\eeqn
On the other hand, since (\ref{eq:3.112-25j}), (\ref{eq:3.112-26j}) lead to 
$\lambda_n^{(1)} > \delta_n \ge \sqrt{\alpha_n}$,
it follows from (\ref{eq:3.112-28j}) that

\beq\label{eq:3.112-29j}
\Pr\left\{\nth\log\frac{W^n(\tilde{Y}^n| \tilde{X}^n)}{P_{Y^n}(\tilde{Y}^n)} \le 
\nth\log\frac{1}{P_{V^n}(\tilde{V}^n)} + \gamma_n \right\} \le 
\sqrt{\alpha_n}.
\eeq
 Then, by the definition of $\tilde{V}^n$,
\[
\nth\log\frac{1}{P_{V^n}(\tilde{V}^n)} \ge d_n,
\]
and so from (\ref{eq:3.112-29j}), we obtain
\beq\label{eq:3.112-30j}
\Pr\left\{\nth\log\frac{W^n(\tilde{Y}^n| \tilde{X}^n)}{P_{Y^n}(\tilde{Y}^n)} \le 
d_n + \gamma_n \right\} \le 
\sqrt{\alpha_n}.
\eeq
Next, since it follows from (\ref{eq:3.112-27j}) that
\beqns
P_{Y^n}(\ssy) & = & 
\lambda_n^{(1)}P_{\tilde{Y}^n}(\ssy) +
 \lambda_n^{(2)}P_{\overline{Y}^n}(\ssy)\\
& \ge & \lambda_n^{(1)}P_{\tilde{Y}^n}(\ssy)\\
& \ge & \delta_n P_{\tilde{Y}^n}(\ssy)\\
& \ge & e^{-n\gamma_n^{\prime}} P_{\tilde{Y}^n}(\ssy),
\eeqns
we have 
\[
\nth\log\frac{1}{P_{Y^n}(\tilde{Y}^n)}\le
\nth\log\frac{1}{P_{\tilde{Y}^n}(\tilde{Y}^n)} + \gamma_n^{\prime},
\]
which is substituted into (\ref{eq:3.112-30j}) to get
\beq\label{eq:3.112-31j}
\Pr\left\{\nth\log\frac{W^n(\tilde{Y}^n| \tilde{X}^n)}{P_{\tilde{Y}^n}(\tilde{Y}^n)} \le 
d_n + \gamma_n -  \gamma_n^{\prime} \right\} \le 
\sqrt{\alpha_n}.
\eeq
On the other hand, by the definition (\ref{eq:3.112-25j}) of $d_n$,
\beq\label{eq:3.112-32j}
\Pr\left\{
\nth\log\frac{1}{P_{V^n}(V^n)} \ge d_n + 2\gamma_n^{\prime} \right\}
\le \delta_n.
\eeq
Set $c_n = d_n +  2\gamma_n^{\prime}$ and note that 
$\alpha_n \to 0,$ $\delta_n \to 0$ $(n\to\infty)$ and 
$ \gamma_n^{\prime} =\frac{\gamma_n}{4}$, then 
by (\ref{eq:3.112-31j}), (\ref{eq:3.112-32j}) we have
\beqns
\lefteqn{\lmn \left(
\Pr\left\{\nth\log\frac{1}{P_{V^n}(V^n)} 
\ge c_n \right\}\right.}\nonumber\\
& & \quad\quad\quad  +  
\left.\Pr\left\{
\nth\log\frac{W^n(\tilde{Y}^n|\tilde{X}^n)}{P_{\tilde{Y}^n}
(\tilde{Y}^n)}
 \le c_n + \frac{1}{4}\gamma_n\right\}\right) 
 = 0. 
\eeqns
Finally, resetting $\tilde{X}^n\tilde{Y}^n$, $ \frac{1}{4}\gamma_n$ as $X^nY^n$ and
$\gamma_n$, respectively, we conclude that  condition 2), i.e., 
 (\ref{eq:3.112-23j}) holds.
\QED

Having  established an  information-spectrum separation-like form of 
 the sufficient condition 
(\ref{eq:3.112-12j}) in Theorem \ref{teiri:3.17-1j}, let us now turn to demonstrate  
several information-spectrum versions derived from the necessary condition 
(\ref{eq:3.112-17j}) in Theorem \ref{teiri:3.17-2j}.

\begin{meidai}[Necessary conditions]\label{mei:dai1}  
{\rm The following two  are 
necessary conditions  for the transmissibility.}
{\rm

\medskip  

1 ) For {\em some} channel input $\sX =\{X^n\}\noi$ and 
for {\em any} sequence $\{\gamma_n \}\noi$ satisfying  condition (\ref{eq:3.112-11j}),
 it holds that
\beq\label{eq:3.112-33js1}
\lmn \Pr\left\{\MID \le 
\nth\log\frac{1}{P_{V^n}(V^n)} - \gamma_n \right\}=0,
\eeq
where $Y^n$ is the channel output via $W^n $ due to  the channel input $X^n$
and $V^n\to X^n\to Y^n$.
}

\medskip

{\rm
2)
For {\em any} 
 sequence $\{\gamma_n \}\noi$ satisfying  condition (\ref{eq:3.112-11j})
 and for  {\em some} channel input $\sX =\{X^n\}\noi$,
 %
it holds that
\beq\label{eq:3.112-33jtr2}
\lmn \Pr\left\{\MID \le 
\nth\log\frac{1}{P_{V^n}(V^n)} - \gamma_n \right\}=0,
\eeq
where $Y^n$ is the channel output via $W^n $ due to  the channel input $X^n$
and  $V^n\to X^n\to Y^n$.
}
\\
\end{meidai}

\noindent
{\em Proof:}\  The necessity of condition 1)  immediately follows from 
necessity condition (\ref{eq:3.112-17j}) in Theorem \ref{teiri:3.17-2j}.
Moreover, it is also trivial to see that condition 1) implies condition 2) as an immediate 
logical consequence, and hence condition 2) is also a necessary condition.
\QED

The necessary condition 1) in  Theorem \ref{teiri:3.17-4j} below  is the same as condition 2)
in  Proposition \ref{mei:dai1}. This is written  here  again in order to
emphasize a pleasing duality between Theorem \ref{teiri:3.17-3j}
and Theorem \ref{teiri:3.17-4j}, which reflects on the duality between
two fundamental Lemmas \ref{hodai:source-channel_Direct} and 
\ref{hodai:source-channel_Converse} .
%
%
%
%
%

\begin{teiri}[Equivalence of necessary conditions]\label{teiri:3.17-4j}
{\rm
The following two conditions are equivalent:
}

\medskip

{\rm
1)  For  {\em any} sequence $\{\gamma_n \}\noi$ satisfying  condition (\ref{eq:3.112-11j})
and
for {\em some} channel input $\sX =\{X^n\}\noi$,
 it holds that
\beq\label{eq:3.112-33j}
\lmn \Pr\left\{\MID \le 
\nth\log\frac{1}{P_{V^n}(V^n)} - \gamma_n \right\}=0,
\eeq
where $Y^n$ is the channel output via $W^n $ due to  the channel input $X^n$
 and $V^n\to X^n\to Y^n$.
}

\medskip

{\rm
2) ({\bf Domination})
For {\em any} 
 sequence $\{\gamma_n \}\noi$ satisfying  condition (\ref{eq:3.112-11j})
 and for  {\em some} channel input $\sX =\{X^n\}\noi$  and 
{\em some} sequence $\{c_n\}\noi$,
 %
it holds that
\beqn
\lefteqn{\lmn \left(
\Pr\left\{\nth\log\frac{1}{P_{V^n}(V^n)} 
\ge c_n \right\}\right.}\nonumber\\
& & \quad\quad\quad  +  
\left.\Pr\left\{\MID \le c_n - \gamma_n\right\}\right) 
 = 0, \label{eq:3.112-34j}
\eeqn
where $Y^n$ is the channel output via $W^n $ due to  the channel input $X^n$.}
\end{teiri}

\noindent
{\em Proof:}\  

This theorem can be proved in the entirely same manner as in the proof of Theorem  \ref{teiri:3.17-3j} with $\gamma_n$ replaced by $-\gamma_n$.
\QED


\bchui\label{chui:3.13-8j}
{\rm
Originally, the definition of {\em domination} given by Vembu, Verd\'u and 
Steinberg \cite{vembu-verdu-stein} is not  condition 2) in Theorem 
\ref{teiri:3.17-4j} but the following:
\\

$2^{\prime}$) (Domination)  For {\em any} sequence $\{d_n\}\noi$ and
 {\em any} sequence $\{\gamma_n\}\noi$ satisfying  condition  
(\ref{eq:3.112-11j}), there exists  {\em some} channel input 
$\sX =\{X^n\}\noi$ such that
\beqn
\lefteqn{\lmn \left(
\Pr\left\{\nth\log\frac{1}{P_{V^n}(V^n)} 
\ge d_n \right\}\right.}\nonumber\\
& & \quad\quad\quad @\times
\left.\Pr\left\{\MID \le d_n - \gamma_n\right\}\right) 
 = 0 \label{eq:3.112-35j}
\eeqn
holds, where $Y^n$ is the channel output via $W^n $ due to  the channel input $X^n$.
 \QED

This necessary condition
$2^{\prime})$ is implied by  necessary condition 2) 
in Theorem \ref{teiri:3.17-4j}.
To see this, set
\beqn\label{eq:Matusi-1}
\alpha_n & \equiv & \Pr\left\{\nth\log\frac{1}{P_{V^n}(V^n)} 
\ge c_n \right\},\label{eq:Matusi-1q}\\
\beta_n &  \equiv & \Pr\left\{\MID \le c_n - \gamma_n\right\},
\label{eq:Matusi-2}\\
\kappa_n & \equiv & \Pr\left\{\nth\log\frac{1}{P_{V^n}(V^n)} 
\ge d_n \right\},\label{eq:Matusi-3}\\
\mu_n &  \equiv & \Pr\left\{\MID \le d_n - \gamma_n\right\}.
\label{eq:Matusi-4}
\eeqn
Then, we observe that $\kappa_n \le \alpha_n$ if $d_n \ge c_n$; and 
$\mu_n \le \beta_n$ if $d_n \le c_n$, and hence  it follows from condition 2) that 
$\kappa_n\mu_n \le \alpha_n + \beta_n$ $\to 0$ as $n$ tends to $\infty$. 
Thus, condition 2) implies condition  $2^{\prime})$, which means that
condition 2) is strictly  stronger than or equivalent to condition $2^{\prime})$ as 
necessary conditions for the transmissibility. It is not currently clear, however, 
whether both are equivalent or not.
%
%
%
%
}
\chuiowari
\echui
\bchui\label{chui:domZ}{\rm
Condition 2) in Theorem \ref{teiri:3.17-4j}  of this form is used later to directly prove 
Theorem \ref{teiri:3.17-5j_M-UTU-1} (separation theorem), while condition $2^{\prime})$
in Remark \ref{chui:3.13-8j} of this form is irrelevant for this purpose. \QED
}
\echui

\section{$\vep$-Transmissibility Theorem}
\label{ss:ep-achiev}

So far we have considered only the case where the error probability $\vep_n$
satisfies the condition ${\dss \lmn \vep_n =0}$. However, we can relax this condition as follows:
\beq\label{3.227-ep-achiv}
 \lsn \vep_n \le \vep,
\eeq
where $\vep$ is any constant such that 
$0 \le \vep <1$. (It is obvious that the special case with $\vep =0$ coincides with the case that we have considered so far.)  We now say that the source $\sV$  is $\vep$-{\em transmissible} over the channel $\sW$ when  there exists an $(n, \vep_n)$ code satisfying 
 condition  (\ref{3.227-ep-achiv}).

Then, the same  arguments as in the previous sections with due slight modifications  lead to the following two theorems in parallel with Theorem \ref{teiri:3.17-1j} and 
Theorem \ref{teiri:3.17-2j}, respectively: 

\begin{teiri}[$\vep$-Direct theorem]\label{teiri:3.17-1j-kruger}
{\rm
Let $\sV =\{V^n\}\noi$, $\sW =\{W^n\}\noi$ be a general source and a general channel, respectively. If there exist  {\em some} channel input 
$\sX =\{X^n\}\noi$  and   {\em some} sequence $\{\gamma_n\}\noi$
such that
\beq\label{eq:3.112-11j_E}
\gamma_n > 0, \ \gamma_n \to 0\ \mbox{and}\ n\gamma_n \to \infty \quad (n\to\infty)
\eeq
for which it holds that 
\beq\label{eq:3.112-12j-kruger}
\lsn \Pr\left\{\MID \le 
\nth\log\frac{1}{P_{V^n}(V^n)} +\gamma_n \right\}\le \vep,
\eeq
then  the source $\sV =\{V^n\}\noi$ is $\vep$-transmissible over the channel 
$\sW =\{W^n\}\noi$, where 
$Y^n$ is the channel output via $W^n $ due to  the channel input $X^n$
 and $V^n\to X^n \to Y^n$.
}
\QED
\end{teiri}

\begin{teiri}[$\vep$-Converse theorem]\label{teiri:3.17-2j-kruger}
{\rm
Suppose that  a general source $\sV =\{V^n\}\noi$ is $\vep$-transmissible 
over a general channel $\sW =\{W^n\}\noi$, %
and let
the channel input  be $\sX = \{X^n \equiv \varphi_n(V^n)\}\noi$
where $\varphi_n: \cV^n\to \cX^n$
is the channel encoder. Then, 
%
 for  {\em any} sequence $\{\gamma_n \}\noi$
satisfying  condition (\ref{eq:3.112-11j_E}), it holds that
\beq\label{eq:3.112-17j-kruger}
\lsn \Pr\left\{\MID \le 
\nth\log\frac{1}{P_{V^n}(V^n)} -\gamma_n \right\}\le \vep,
\eeq
where $Y^n$ is the channel output via $W^n $ due to  the channel input $X^n$
and $V^n\to X^n \to Y^n$.
}
\QED
\end{teiri}

\bchui\label{chui:daini10}
{\rm 
It should be noted here that such a sufficient condition 
(\ref{eq:3.112-12j-kruger}) as well as such a
necessary condition (\ref{eq:3.112-17j-kruger}) for the $\vep$-transmissibility 
cannot actually be derived in the way of generalizing the strict domination in 
(\ref{eq:3.112-23j}) and the domination in (\ref{eq:3.112-34j}).
It should be noted also that, under the $\vep$-transmissibility criterion, 
joint source-channel coding is beyond the separation principle.
\QED
}\echui

\section{Separation Theorems of the Traditional Type
\label{ss:SEpThTra-1}}

Thus far we have investigated the joint source-channel coding
problem from the viewpoint of information spectra and established the fundamental
theorems (Theorems \ref{teiri:3.17-1j}$\sim$\ref{teiri:3.17-4j}).
These results are of seemingly  different forms 
from separation theorems of the traditional type. 
Then, it would be natural to ask a question how  
the separation principle of the information spectrum type is related to 
 separation theorems of the traditional type. In this section we address
this question. 

To do so, we first need some preparation. We denote by $R_f(\sV)$ the infimum
of all achievable fixed-length coding rates for a general source $\sV=\{V^n\}\noi$
(as for the formal definition, see  Han and Verd\'u \cite{han-verdu-approx},  Han \cite[Definitions 1.1.1, 1.1.2]{Ha98}), and 
denote by $C(\sW)$ the capacity of a general channel $\sW=\{W^n: \cX^n \to \cY^n\}\noi$
(as for the formal definition, see Han and Verd\'u \cite{han-verdu-approx},  Han 
\cite [Definitions 3.1.1, 3.1.2]{Ha98}). First,  $R_f(\sV)$  is characterized as
\medskip 

\begin{teiri}[{\rm Han and Verd\'{u} \cite{han-verdu-approx}, Han\cite{Ha98}}] 
\label{teiri:Han_Verdu}
{\rm 
\mbox{}
\beq\label{eq:folk-5}
R_f(\spr{V})= \barH(\spr{V}),
\eeq
where
}
\footnote{For an arbitrary sequence of real-valued random variables 
$\{ Z_n \} _{n=1}^{\infty}$, we define the following notions (cf. Han and Verd\'u \cite{han-verdu-approx},
Han\cite{Ha98}):
$\plimsup_{n \to \infty} Z_n $  $\equiv $ $\inf \{ \alpha \mid \lim_{n \to \infty} $ 
$\Prob{Z_n > \alpha} = 0 \}$ (the {\em limit superior in probability}), and
$\pliminf_{n \to \infty} Z_n $  $\equiv $ $\sup \{ \beta \mid \lim_{n \to \infty} $ 
$\Prob{Z_n< \beta} = 0  \}$ (the {\em limit inferior in probability}). 
}

\beq\label{eq:folk-6}
\barH(\spr{V}) = \mbox{{\rm p-}}\limsup_{n \to \infty} \frac{1}{n} \log \frac{1}{P_{V^n}(V^n)}.
\eeq
\end{teiri}

\medskip
Next, let us consider about the characterization of $C(\sW)$. Given a general channel $\sW=\{W^n\}\noi$ and its input  $\sX=\{X^n\}\noi$, let  
$\sY=\{Y^n\}\noi$ be the output due to the input $\sX=\{X^n\}\noi$ via the channel
$\sW=\{W^n\}\noi$. Define
\bteigi \label{ubarI}
\mbox{}
\beq\label{donma2}
\uI(\sX;\sY) =
\mbox{{\rm p-}}\liminf_{\nti} \nth \log \frac{W^n(Y^n|X^n)}{P_{Y^n}(Y^n)}.
\eeq
\eteigi
Then,  the capacity $C(\sW)$ is characterized as follows.
\bteiri[{\rm Verd\'{u} and
 Han \cite{verdu-han-capacity}, Han\cite{Ha98}}] \label{teiri:gen_chan}
 {\rm 
\beq \label{eq:general-capacity-formula}
C(\sW) = \sup_{\sX} \uIXY,
\eeq
where $ \sup_{\sX}$ means the supremum over all possible inputs $\sX$. \QED
}
\eteiri

With these preparations, let us turn to the separation theorem problem 
of the traditional type.
A general source $\sV=\{V^n\}\noi$ is said to be {\em information-stable}
 (cf. Dobrushin \cite{dobrushin}, Pinsker \cite{pinsker}) if
\beq\label{eq:Mujo-1}
\frac{\nth\log \frac{1}{P_{V^n}(V^n)}}{H_n(V^n)} \to 1 \quad 
\mbox{in prob.},
\eeq
where $H_n(V^n)=\nth H(V^n)$ and $H(V^n)$ stands for the entropy of $V^n$
 (cf. Cover and Thomas \cite{cover-thomas}).
Moreover, a general channel  $\sW=\{W^n\}\noi$ is said to be 
{\em information-stable} 
(cf. Dobrushin \cite{dobrushin}, Pinsker \cite{pinsker}, Hu \cite{huguoding}) 
if there exists a channel input $\sX=\{X^n\}\noi$ such that
\beq\label{eq:Mujo-2}
\frac{\nth\log \frac{W^(Y^n|X^n)}{P_{Y^n}(Y^n)}}{C_n(W^n)} \to 1 \quad \mbox{in prob.},
\eeq
where 
\[
C_n(W^n) = \sup_{X^n}\nth I(X^n;Y^n),
\]
and $Y^n$ is the channel output via $W^n$ due to the channel input $X^n$;
 and $I(X^n;Y^n)$ is the mutual information between $X^n$ and $Y^n$
(cf. Cover and Thomas \cite{cover-thomas}). 
Then, we can summarize a typical  separation theorem of the traditional type
as follows.
\begin{teiri}[{\rm Dobrushin \cite{dobrushin}, Pinsker \cite{pinsker}}]
\label{teiri:3.17-5j_M} 
{\rm Let  the channel 
$\sW=\{W^n\}\noi$ be information-stable and  suppose that the limit
${\dss \lmn C_n(W^n)}$ exists, or, let the source $\sV=\{V^n\}\noi$  be 
information-stable and suppose that the limit ${\dss \lmn H_n(V^n)}$ exists.
Then, the following two statements hold:
\begin{enumerate}
\renewcommand{\labelenumi}{\arabic{enumi})}
\item 
 If $R_f(\sV) < C(\sW)$, then the source $\sV$ is transmissible 
over the channel
$\sW$. In this case, we can separate 
the source coding and the channel coding.
\item 
If the source $\sV$ is transmissible over the channel
$\sW$, then it must hold that $R_f(\sV) \le C(\sW)$. \QED
\end{enumerate}
}
\end{teiri}

In order to generalize Theorem~\ref{teiri:3.17-5j_M}, we need 
to introduce the concept of {\it optimistic}
 coding. The ``optimistic'' standpoint means
 that we evaluate the coding reliability with error probability $\liminf_{n\to\infty}\vep_n =0$
(that is, $\vep_n <\forall\vep $ \ {\it for infinitely many} $n$).
In contrast  with this, the standpoint that we have taken so far is called 
{\it pessimistic}\index{pessimistic@pessimistic coding} 
with error probability $\lim_{n\to\infty}\vep_n =0$ (that is, 
$\vep_n <\forall\vep $ \ {\it for all sufficiently large} $n$).

The following one concerns the optimistic source coding with 
any general source $\sV$.
\begin{teigi}[Optimistic achievability for source coding]\label{teigi:ach_FFcode-Springer}
\mbox{}
{\rm 
\begin{eqnarray*}
\mbox{\rm Rate $R$ is optimistically achievable} & \defarrow & 
\mbox{\rm There exists an $(n,M_n,\varepsilon_n)$- source code} \\
& &  \mbox{\rm satisfying ${\displaystyle \liminf_{n \rightarrow
\infty}}\varepsilon_n = 0$ and} \\ & & \mbox{\rm 
 ${\displaystyle \limsup_{n \rightarrow \infty} 
\frac{1}{n} \log M_n \le R}$},
\end{eqnarray*}
where $\nth \log M_n$ is the coding rate per source letter (see, e.g., Han \cite [Section 1.1] {Ha98}).
}
\end{teigi}

\begin{teigi}[\emp{Optimistic  achievable fixed-length coding rate}]
\index{infimum achievable fixed length coding rate@
infimum achievable fixed-length coding coding rate} \label{teigi:cap_FFcode-Springer}
\mbox{}
\[
\underline{R}_f(\spr{V})= \inf \left\{ R \mid \mbox{\rm $R$ is
optimistically achievable}
\right\}.
\]
\end{teigi}

\bigskip
\noindent
Then, for any general source $\sV=\{V^n\}\noi$ we have: 
\begin{teiri}[{\rm Chen and Alajaji \cite{chen-alajaji}}] 
\label{teiri:3.17-5j_M-SPringer} 
\beq\label{eq:SpriB-1}
\underline{R}_f(\sV) = \inf\left\{ R \left| \liminf_{n\to\infty}
\Pr\left\{\nth \log \frac{1}{P_{V^n}(V^n)}\ge R\right\}
=0\right.\right\}. 
\eeq
\end{teiri}
%

On the other hand, the next one concerns the optimistic channel capacity.
\begin{teigi}[Optimistic achievability for channel coding]\label{teigi:ach_VFcode-Springer-2}
\mbox{}
\begin{eqnarray*}
\mbox{\rm Rate $R$ is optimistically achievable} & \defarrow & 
\mbox{\rm There exists an $(n,M_n,\varepsilon_n)$-channel code} \\
& &  \mbox{\rm satisfying ${\displaystyle \liminf_{n \rightarrow
\infty}}\varepsilon_n = 0$ and} \\ & & \mbox{\rm 
 ${\displaystyle \liminf_{n \rightarrow
\infty} 
\frac{1}{n} \log M_n \ge R}$,}
\end{eqnarray*}
{\rm 
where $\nth \log M_n$ is the coding rate per channel use (see, e.g., Han \cite [Section 3.1]{Ha98}).
}
\end{teigi}

\begin{teigi}[\emp{Optimistic channel capacity}]
\index{infimum achievable fixed length coding rate@
infimum achievable fixed-length coding rate} \label{teigi:cap_VFcode-Springer-2}
\mbox{}
\[
\overline{C}(\sW)= \sup \left\{ R \mid \mbox{\rm $R$ is
optimistically achievable}
\right\}.
\]
\end{teigi}

\bigskip
\noindent
Then, with a general channel $\sW=\{W^n\}\noi$ we have
\begin{teiri}[{\rm 
Chen and Alajaji \cite{chen-alajaji}}] 
\rm
\label{teiri:3.17-5j_M-SPringer-Spr-4} 
\beqn\label{eq:SpriB-5}
\lefteqn{\overline{C}(\sW) }\nonumber\\
&=&
 \sup_{\sX}
\sup\left\{ R \left| \liminf_{n\to\infty}
\Pr\left\{\nth \log \frac{W^n(Y^n|X^n)}{P_{Y^n}(Y^n)}\le R\right\}
=0\right.\right\}, 
\eeqn
where $Y^n$ is the output due to the input $\sX=\{X^n\}\noi$.
\QED
\end{teiri}

\bchui\label{chui:Utik-1}
{\rm
It is not difficult to check that, in parallel with Theorem~\ref{teiri:3.17-5j_M-SPringer} and 
Theorem
\ref{teiri:3.17-5j_M-SPringer-Spr-4}, Theorem~\ref{teiri:Han_Verdu} 
and Theorem~\ref{teiri:gen_chan}
can be rewritten as
\beqn
R_f(\sV) & = & \inf\left\{ R \left| \lim_{n\to\infty}
\Pr\left\{\nth \log \frac{1}{P_{V^n}(V^n)}\ge R\right\}
=0\right.\right\}, \label{eq:inoyaka-1}\\
C(\sW) & = &
 \sup_{\sX}
\sup\left\{ R \left| \lim_{n\to\infty}
\Pr\left\{\nth \log \frac{W^n(Y^n|X^n)}{P_{Y^n}(Y^n)}\le R\right\}
=0\right.\right\},\nonumber\\
\label{eq:inoyaka-2}
\eeqn
from which, together with Theorem~\ref{teiri:3.17-5j_M-SPringer} and 
Theorem
\ref{teiri:3.17-5j_M-SPringer-Spr-4}, it immediately follows that
\beqn
C(\sW) & \le & \overline{C}(\sW), \label{eq:UTUI-1} \\
\underline{R}_f(\sV) & \le & R_f(\sV). \label{eq:UTUI-2}
\eeqn
}
\echui

\bigskip
\noindent
Now, we have: 
\begin{teiri}
\label{teiri:3.17-5j_M-UTU-1} 
{\rm 
Let   
$\sW=\{W^n\}\noi$ be a general channel and  $\sV=\{V^n\}\noi$  be a general
source.  Then, the following two statements hold:
\begin{enumerate}
\renewcommand{\labelenumi}{\arabic{enumi})}

\item 
If $R_f(\sV) < C(\sW)$, then the source $\sV$ is transmissible over the channel
$\sW$. In this case, we can separate the source coding and 
the channel coding.

\item 
If the source $\sV$ is transmissible over the channel
$\sW$, then it must hold that 
\beqn
\underline{R}_f(\sV) &\le & C(\sW),\label{eq:UTUI-3}\\
R_f(\sV) &\le & \overline{C}(\sW).\label{eq:UTUI-4}
\eeqn

\end{enumerate}
}
\end{teiri}
\bchui\label{chui:Utik-3}
{\rm
As was mentioned in Remark \ref{chui:domZ}, 
we use Theorem \ref{teiri:3.17-4j} in order to prove (\ref{eq:UTUI-3}) and (\ref{eq:UTUI-4}), where 
inequality (\ref{eq:UTUI-4}) was shown in a rather roundabout manner  by  Vembu, Verd\'u and  Steinberg \cite{vembu-verdu-stein} (invoking Domination 
$2^{\prime}$) in Remark \ref{chui:3.13-8j} instead of Domination 2) in Theorem
 \ref{teiri:3.17-4j}).
 \QED
}
\echui
\noindent
{\it Proof of Theorem~\ref{teiri:3.17-5j_M-UTU-1}}.

\noindent
1): Since 
$R_f(\sV) = \overline{H}(\sV)$, $ C(\sW) = \sup_{\sX}\underline{I}(\sX; \sY)$ 
 by Theorem~\ref{teiri:Han_Verdu} and Theorem~\ref{teiri:gen_chan}, 
the inequality $ R_f(\sV) < C(\sW)$ implies that condition 2) in 
Theorem~\ref{teiri:3.17-3j} holds for $\sX=\{X^n\}\noi$ 
attaining the supremum $ \sup_{\sX}\underline{I}(\sX; \sY)$ with, 
for example, $c_n=$ $ \frac{1}{2}(R_f(\sV) + C(\sW))$.
Therefore, the source $\sV$ is transmissible over the channel $\sW$.\\

\noindent
2):  If the source $\sV$ is transmissible over the channel
$\sW$, then  condition 2) in Theorem~\ref{teiri:3.17-4j} holds
with some $\{c_n\}\noi$, i.e.,
\beq\label{eq:3.112-36j_UTU-1}
\lmn \Pr\left\{\nth\log\frac{1}{P_{V^n}(V^n)} \ge c_n\right\} = 0,
\eeq
\beq\label{eq:3.112-36j_UTU-2}
\lmn\Pr\left\{\MID \le c_n - \gamma_n\right\}
 = 0.
\eeq
Since $\lmn \gamma_n =0$,
these two conditions with any small constant $\delta >0$ lead us 
to the following formulas:
\beqn
\lin \Pr\left\{\nth\log\frac{1}{P_{V^n}(V^n)} \ge \lin c_n +\delta\right\} & = &0,
\label{eq:3.112-36j_UTU-4}\\
\lmn \Pr\left\{\nth\log\frac{1}{P_{V^n}(V^n)} \ge \lsn c_n +\delta\right\} & = &0,
\label{eq:3.112-36j_UTU-5}
\eeqn
\beqn
\lmn\Pr\left\{\MID \le \lin c_n -\delta  \right\} & =  & 0,
\label{eq:3.112-36j_UTU-7}\\
\lin\Pr\left\{\MID \le \lsn c_n-\delta\right\} & =  & 0.
\label{eq:3.112-36j_UTU-8}
\eeqn
Then, Theorem~\ref{teiri:3.17-5j_M-SPringer} and 
(\ref{eq:3.112-36j_UTU-4}) imply
that $\underline{R}_f(\sV) \le  \lin c_n$, whereas (\ref{eq:3.112-36j_UTU-7}) 
implies that $\underline{I}(\sX; \sY) \ge \lin c_n$. Therefore, by Theorem
\ref{teiri:gen_chan} we have 
\[
\underline{R}_f(\sV) \le  
 \lin c_n \le 
\underline{I}(\sX; \sY) \le \sup_{\sX} \underline{I}(\sX; \sY)= C(\sW).
\]
On the other hand, (\ref{eq:3.112-36j_UTU-5}) implies that 
$\overline{H}(\sV) \le \lsn c_n$. Furthermore, 
(\ref{eq:3.112-36j_UTU-8}) together with 
Theorem~\ref{teiri:3.17-5j_M-SPringer-Spr-4} gives us 
\[
\overline{H}(\sV) \le \lsn c_n \le \overline{C}(\sW).
\] 
Finally, note that $R_f(\sV) = \overline{H}(\sV)$ by Theorem~\ref{teiri:Han_Verdu}.
\QED

We are now interested in the problem of  what conditions
are needed to attain  equalities $\underline{R}_f(\sV) = R_f(\sV)$
and/or $\overline{C}(\sW) = C(\sW)$ in Theorem~\ref{teiri:3.17-5j_M-UTU-1} and so on. To see this, we need the following four definitions: 
\bteigi\label{teigi-UTUka-1s}
{\rm 
A general source $\sV=\{V^n\}\noi$ is said to satisfy the 
{\em  strong converse property} if 
\[
\overline{H}(\sV) = \underline{H}(\sV)
\]
holds (as for the operational meaning, refer to Han \cite{Ha98}), where
}
\[
\underline{H}(\sV) = \mbox{{\rm p-}}
\liminf_{n \to \infty} \frac{1}{n} \log \frac{1}{P_{V^n}(V^n)}.
\]
\eteigi
\bteigi\label{teigi-UTUka-2k}
{\rm 
A general channel $\sW=\{W^n\}\noi$ is said to satisfy the {\em strong converse property}
if 
\beq\label{eq: OYA-2a} 
\sup_{\sbold{X}}\underline{I}(\sX;\sY) =
\sup_{\sbold{X}}\overline{I}(\sX;\sY)
\eeq
holds (as for the operational meaning, refer to Han \cite{Ha98}, 
Verd\'u and Han \cite{verdu-han-capacity}), where
} 
\[
\overline{I}(\sX;\sY) =
\mbox{{\rm p-}}\limsup_{\nti} \nth \log \frac{W^n(Y^n|X^n)}{P_{Y^n}(Y^n)}.
\]
\eteigi
\bteigi\label{teigi-UTUka-1}
{\rm 
A general source $\sV=\{V^n\}\noi$ is said to satisfy the 
{\em  semi-strong converse property} if for {\em all} divergent 
subsequences $\{n_i\}\noi$ of
positive integers such that $n_1<n_2<\cdots \to \infty$ it holds that }
\beq\label{eq: OYA-1}
\mbox{{\rm p-}}\limsup_{i\to\infty}
\frac{1}{n_i}\log\frac{1}{P_{V^{n_i}}(V^{n_i})}
= \overline{H}(\sV).
\eeq
\eteigi
\bteigi\label{teigi-UTUka-2we}
{\rm 
A general channel $\sW=\{W^n\}\noi$ is said to satisfy the {\em semi-strong converse property}
if  for {\em all} divergent 
subsequences $\{n_i\}\noi$ of
positive integers such that $n_1<n_2<\cdots \to \infty$ it holds that }
\beq\label{eq: OYA-2} 
\mbox{{\rm  p-}}\liminf_{i\to\infty}
\frac{1}{n_i}\log\frac{W^{n_i}(Y^{n_i}|X^{n_i})}
{P_{Y^{n_i}}(Y^{n_i})} \le \sup_{\sbold{X}}\underline{I}(\sX;\sY),
\eeq
{\rm where $Y^n$ is the channel output via $W^n$ due to the channel input $X^n$.} \QED
\eteigi
With these definitions, we have the following lemmas:
\bhodai\label{mikenten1} \mbox{}
{\rm 
\begin{itemize}
\item[{1)}] 
The information-stability of a source $\sV$ (resp. a channel $\sW$) 
with the limit implies the strong converse property of $\sV$ (resp. $\sW$). 
 
\item[{2)}] 
The strong converse property of a source $\sV$ (resp. a channel $\sW$) 
implies the semi-strong converse property of $\sV$ (resp. $\sW$). 
 \QED
\end{itemize}
}
\ehodai
\bhodai\label{hodai:qqq-1}
\mbox{}
{\rm 
\begin{enumerate} 
\item[{1)}] A general source $\sV$ satisfies 
the semi-strong converse property if and only if
\beq\label{eq:WWW-1}
\underline{R}_f(\sV)=R_f(\sV). 
\eeq
\item[{2)}] A general channel $\sW$ satisfies 
the semi-strong converse property if and only if 
\beq\label{eq:WWW-2}
\overline{C}(\sW) = C(\sW). 
\eeq
\end{enumerate}
}
\ehodai

\proof 
It is obvious in view of Theorem~\ref{teiri:3.17-5j_M-SPringer}, 
Theorem~\ref{teiri:3.17-5j_M-SPringer-Spr-4} and
Remark \ref{chui:Utik-1}. \QED

\bchui\label{re-houjin}
{\rm 
An operational equivalent of the notion of semi-strong converse property is found in Vembu, Verd\'u and Steinberg \cite{vembu-verdu-stein}.
Originally, Csisz\'ar and K\"orner \cite{csiszar-korner} posed 
 two operational standpoints
in source coding and channel coding, i.e., 
the {\em pessimistic standpoint} and the {\em optimistic standpoint}. 
In their terminology, Lemma~\ref{hodai:qqq-1} states that,
for  source coding, 
the semi-strong convserse property is equivalent to the statement that 
both the pessimistic standpoint and the optiimistic standpoint result 
in the same infimum of all achievable fixed-length source coding rates;
similarly, for  channel coding, 
the semi-strong convserse property
 is equivalent to the claim that 
both the pessimistic standpoint and the optimistic standpoint 
result in the same  supremum of all achievable channel coding rates.}
\QED
\echui

Thus, Theorem~\ref{teiri:3.17-5j_M-UTU-1} together with 
Lemma~\ref{hodai:qqq-1} immediately yields the following stronger 
separation theorem of the traditional type:
\begin{teiri}
\label{teiri:3.17-5j_MW}
{\rm
Let either a general source $\sV=\{V^n\}\noi$ or a general 
channel $\sW=\{W^n\}\noi$ 
satisfy the semi-strong converse property. Then, the following 
two statements hold: 
\begin{enumerate}
\renewcommand{\labelenumi}{\arabic{enumi})}

\item 
 If $R_f(\sV) < C(\sW)$, then the source $\sV$ is transmissible 
over the channel
$\sW$. In this case, we can separate 
the source coding and the channel coding.
\item 
If the source $\sV$ is transmissible over the channel
$\sW$, then it must hold that $R_f(\sV) \le C(\sW)$. \QED

\end{enumerate}
}
\end{teiri}
%
%
%
%
%
%

%
%
\brei\label{rei:asakara1}
{\rm
Theorem \ref{teiri:3.17-5j_M} is an immediate consequence of
Theorem \ref{teiri:3.17-5j_MW} together with Lemma \ref{mikenten1}.
\QED}
\erei

\brei\label{rei:mixed-end-af}
{\rm
Let us consider two different stationary memoryless sources
$\sV_1=\{V^n_1\}\noi$, $\sV_2=\{V^n_2\}\noi$ with countably infinite source alphabet $\cV$, and define its {\em mixed} source 
$\sV=\{V^n\}\noi$ by
\[
P_{V^n}(\ssv) = \alpha_1P_{V^n_1}(\ssv)
+ \alpha_2P_{V^n_2}(\ssv) \quad (\ssv \in \cV^n),
\]
where $\alpha_1$, $\alpha_2$ are positive constants such that
$\alpha_1+\alpha_2 =  1$. Then, this mixed source $\sV=\{V^n\}\noi$
satisfies the semi-strong converse property but
neither the strong converse property nor the information-stability.
}

{\rm
Similarly, let us consider two different stationary memoryless channels
$\sW_1=\{W^n_1\}\noi$, $\sW_2=\{W^n_2\}\noi$ with arbitrary abstract  input and output  alphabets $\cX, \cY$, and define its {\em mixed} channel 
$\sW=\{W^n\}\noi$ by
\[
W^n(\ssy|\ssx) = \alpha_1W^n_1(\ssy|\ssx)
+ \alpha_2W^n_2(\ssy|\ssx) \quad (\ssx \in \cX^n, \ssy \in \cY^n).
\]
 Then, this mixed channel $\sW=\{W^n\}\noi$
satisfies the semi-strong converse property but
neither the strong converse property nor the information-stability.

Thus, in these mixed cases the separation theorem holds.
}
\QED
\erei


\begin{thebibliography}{999}


\bibitem{shannon-1948} C. E. Shannon, ``A mathematical theory of communication," {\em Bell System Technical Journal}, vol.27, pp.379-423, 
pp. 623-656, 1948


\bibitem{feinstein} A. Feinstein, ``A new basic theorem of information theory," {\em IRE Trans. PGIT}, vol.4, pp.2-22, 1954

\bibitem{ashtein} R.B. Ash, {\em Information Theory},
Interscience Publishers, New York, 1965


\bibitem{dobrushin} R. L. Dobrushin, ``A general formulation of the fundamental Shannon theorem in information theory,"  {\em Uspehi Mat. Acad. Nauk. SSSR}, vol.40, pp.3-104, 1959: Translation in {\em Transactions of American Mathematical Society}, Series 2, vol.33, pp.323-438, 1963


\bibitem{pinsker} M. S. Pinsker, {\em Information and Information Stability of Random Variables and Processes},  Holden-Day, San Francisco, 1964

\bibitem{huguoding} G. D. Hu,
``On Shannon theorem and its converse for sequence of communication schemes in the case of abstract random variables," 
in {\em  Trans. 3rd Prague Conference on Information Theory, Statistical Decision Functions, Random Processes}, Czechslovak Academy of Sciences,
Prague, pp. 285-333, 1964

\bibitem{han-verdu-approx} T.S. Han and S. Verd\'u, ``Approximation theory of output statistics,"  {\em  IEEE Transactions on Information Theory}, vol.IT-39, no.3, pp. 752-772, 1993

\bibitem{verdu-han-capacity}  S. Verd\'u and T.S. Han, ``A general formula for channel capacity," {\em IEEE Transactions on Information Theory}, vol.IT-40, no.4, pp.1147-1157, 1994


\bibitem{vembu-verdu-stein} S. Vembu, S. Verd\'u and Y. Steinberg,
``The source-channel separation theorem revisited," 
{\em  IEEE Transactions on Information Theory}, vol.IT-41, no.1, 
pp. 44-54, 1995


\bibitem{verdu-han-aep} S. Verd\'u and T. S. Han,
``The role of the asymptotic equipartition property
in noiseless source coding,"  {\em  IEEE Transactions on Information Theory}, vol.IT-43, no.3, pp.847-857, 1997


\bibitem{Ha98} T. S. Han, {\it Information-Spectrum Methods in Information 
Theory}, 
Springer Verlag, New York, 2003



\bibitem{csiszar-korner} I. Csisz\'ar and J. K\"orner, {\em 
Information Theory: Coding Theorems  for Discrete Memoryless Systems}, 
 Academic Press, New York, 1981

\bibitem{cover-thomas} T. M. Cover and J. Thomas, {\em Elements of 
Information Theory},  Wiley, New York, 1991

\bibitem{chen-alajaji} P.N. Chen and F. Alajaji, ''Optimistic Shannon coding theorems
for arbitrary single-user systems," {\em  IEEE Transactions on 
Information Theory},  IT-45, pp. 2623-2629, 1999

\end{thebibliography}
\end{document}